\documentclass[aps,prb,superscriptaddress,showpacs,floatfix,twocolumn]{revtex4-1}

\usepackage{times}
\usepackage{graphicx,graphics,color}
\usepackage{amssymb}
\usepackage{bm}
\usepackage{amsmath}
\usepackage{gensymb}
\usepackage{caption3}
\usepackage{subfigure}
\usepackage{ulem}

\newcommand{\Zn}	{Ba(Fe$_{1-x}$Zn$_{x}$)$_{2}$As$_{2}$\,}
\newcommand{\BFA}	{BaFe$_{2}$As$_{2}$\,}

\begin{document}

\title{Effects of single- and multi-substituted Zn ions in doped-122 type iron-based superconductors}

\author{YuanYuan Zhao}
\affiliation{Department of Physics and Texas Center for Superconductivity, University of Houston, Houston, Texas 77204, USA}

\author{Bo Li}
\affiliation{Department of Physics and Texas Center for Superconductivity, University of Houston, Houston, Texas 77204, USA}

\author{Wei Li}
\affiliation{State Key Laboratory of Functional Materials for Informatics and Shanghai Center for Superconductivity, Shanghai Institute of Microsystem and Information Technology, and CAS-Shanghai Science Research Center, Chinese Academy of Sciences, Shanghai 200050, China}

\author{Hong-Yi Chen}
\affiliation{Department of Physics and Texas Center for Superconductivity, University of Houston, Houston, Texas 77204, USA}
\affiliation{National Taiwan Normal University, Department of Physics, Taipei 116, Taiwan}

\author{Kevin E. Bassler}
\affiliation{Department of Physics and Texas Center for Superconductivity, University of Houston, Houston, Texas 77204, USA}

\author{C. S. Ting}
\affiliation{Department of Physics and Texas Center for Superconductivity, University of Houston, Houston, Texas 77204, USA}
\date{\today}

\begin{abstract}
Recent experiments on Zn-substituted 122-type iron-based superconductors (FeSCs) at electron- and hole- doped region provide us with a testing ground for understanding the effect of Zn impurities in these systems. Our first-principle calculations of the electronic structure reveal that the Zn $3d$ orbitals are far below the Fermi level and chemically inactive, while the Zn $4s$-orbital is partially occupied and its wave function overlapping with those $3d$-orbitals of neighboring Fe-ions. This suggests that the impurity effect is originating in the Zn $4s$-orbital, not its $3d$-orbitals. Employing a phenomenological two-orbital lattice model for 122-FeSCs and the self-consistent Bogoliubov-de Gennes equations, we study how the Zn-impurities suppress the superconductivity in electron- and hole- doped compounds. Our obtained results qualitatively agree with the experimental measurements.
\end{abstract}

\maketitle
\section{Introduction}
In iron-based superconductors (FeSCs), doping can be made by partial substitution of Co and Ni for Fe~\cite{Sefat2008, Li2009}, or substitution of K and Na for Ba~\cite{Rotter2008L, Avci2014} in the antiferromagnetic (AFM) parent compounds. With doping increasing, the spin-density-wave (SDW) order is suppressed and superconductivity (SC) emerges. In addition, the nonmagnetic impurity could be regarded as an important probe in understanding pairing symmetry in superconductors. According to Anderson's theorem~\cite{Anderson1959, Abrikosov1994}, nonmagnetic impurities may not cause pair-breaking in conventional $s$-wave superconductors, but they severely suppress SC transition temperature $T_{c}$ in $d$-wave~\cite{Tarascon1987} and $s_{\pm}$-wave~\cite{Bang2009, Onari2009} superconductors. Studying the effect due to nonmagnetic impurities in FeSCs becomes an indispensable avenue to understand the superconducting physics in these compounds.

In 122-FeSCs, Zn substitution for the Fe ion is preferred as an ideal nonmagnetic impurity~\cite{Yao2012, Li2013, HChen2013, Pan2014}. Most recently, Li \textit{et al.} report~\cite{Li2015} that SC can hardly survive with $3\%$ Zn substitutions in the hole-doped Ba$_{0.5}$K$_{0.5}$Fe$_{2}$As$_{2}$. They also show that the local destruction of SC may provide evidence for $s$-wave pairing symmetry. While the measurements~\cite{Li2011B, Li2012} in the electron-doped BaFe$_{1.89-2x}$Zn$_{2x}$Co$_{0.11}$As$_{2}$ demonstrate that the SC is completely suppressed above a concentration of roughly $8\%$ Zn, regardless of whether the sample is in under-, optimal-, or over- doped regimes. On the other hand, the experimentally observed $T_{c}$ suppression is much slower than those predicted by the theory for the $s_{\pm}$-wave pairing state~\cite{Onari2009}. A recent work~\cite{HChen2013} by Chen \textit{et al.} appears to be able to account for the suppression of SC at roughly $8\%$ Zn in optimally electron-doped 122-FeSCs. So far there exists no theory which is able to consistently explain the different $T_{c}$ suppressions for the electron- and the hole- doped 122-FeSCs. In order to understand this difference, we reexamine the nature of Zn impurities by the first-principles calculations and construct a model for describing the substituted-Zn in \BFA.

Zn element has a $3d^{10}4s^{2}$ electronic configuration. Generally it shows divalent in the compounds. Similar to earlier studies~\cite{Zhang2009, Wadati2010, Berlijn2012, Ideta2013}, our first-principles calculations~\cite{footnote}, as shown in Fig.~\ref{pDOS}, demonstrate that Zn-$3d$ states are far below Fermi energy by about $8$ eV. In addition, our calculations demonstrate that the peaks of Zn-$4s$ level narrowly distribute above and below Fermi energy. More importantly, the empty Zn-$4s$ level is not far above the Fermi energy, which suggests Zn-4$s$ orbitals are partially occupied and Zn has not a valence of two. Thus the heterovalent doping effect of substituted-Zn should be considered in the system.

The conventional way of treating substituted-Zn~\cite{HChen2013, Pan2014} in the mean-field frame is to assume that there exists overlapping between Zn $3d$ and its neighboring Fe $3d$ orbitals. Meanwhile, in order to reflect that the Zn-$3d$ level is far below the Fermi energy, a strongly negative potential-scattering-term is assigned at the Zn site. According to our first-principles calculations, the fully occupied $3d$ orbitals of the substituted-Zn impurity should be regarded strongly localized, and do not contribute to the electron density of states near the Fermi energy. While the partially occupied $4s$ orbital of the substituted-Zn should be the only responsible orbital for the impurity scattering. Instead of considering Zn $3d$-orbitals, we study the effect due to its $4s$-orbital. Because the energy of the empty Zn-$4s$ level is close to the Fermi energy, it should be reasonable to choose a scattering potential of intermediate strength at Zn site.

In this paper, we use an effective two-orbital tight-binding model~\cite{DGZhang2009, YuanYenepl} to describe the \BFA system without substituted-Zn. For the substituted-Zn, there is only one $4s$ orbital, we have to adjust the hopping-parameters between Zn and adjacent Fe sites. These parameters are chosen to fit the experimental results for the optimally electron-doped 122-FeSCs~\cite{Li2011B, Li2012}. Then we employ the fixed parameters to study the $T_{c}$ suppressions for the electron-doped 122-FeSCs at the under-doped region and for the hole-doped 122-FeSCs. Employing self-consistent lattice Bogoliubov-de Gennes (BdG) equations, we demonstrate that our obtained results are qualitatively comparable with those experiments~\cite{Li2011B, Li2012, Li2015} on $T_{c}$ suppressions in various 122-FeSCs with substituted-Zn.

\section{Model Construction}
In the following, we discuss the details about our model for performing the calculations. In the parent compound \BFA, the Fe ions form a square lattice, while the As anions sit alternatively below and above the center of each plaquette of the Fe lattice. This structure contains two intertwined sublattices of Fe ions denoted by $A$ and $B$. Heretofore there are several microscopic multi-orbital models describing iron-based superconductors~\cite{Raghu2008, Kuroki2009, Laad2009, YuanYenepl, DGZhang2009, Wang2010, Thomale2011, Su2012, HuS4}. We choose an effective model~\cite{YuanYenepl}, which has been tested by capturing several important features of the \BFA compounds in good agreement with experiments. This two-orbital tight-binding model takes $A$ and $B$ two Fe ions per unit-cell including Fe-3$d_{xz}$ and Fe-3$d_{yz}$ orbitals. It has also been proven that this model can represent within one Fe atom per unit cell after a gauge transformation~\cite{HChen2013}. The full Hamiltonian of the \BFA system could be written as
\begin{equation}\label{HamTot}
    H = H_{BCS} + H_{int} + H_{imp}
\end{equation}
Here, $H_{BCS}$ is the BCS-type Hamiltonian, including the hopping term and the pairing term, expressed as
\begin{equation}
\begin{split}\label{HamBCS}
    H_{BCS} = &\sum_{i, j, \alpha, \beta, \sigma} t_{i\,j}^{\alpha\, \beta} c_{i\, \alpha\, \sigma}^{\dag} c_{j\, \beta\, \sigma} - \sum_{i, \alpha, \sigma} \mu\, c_{i\, \alpha \, \sigma}^{\dag} c_{i\, \alpha \, \sigma} \\
    &+ \sum_{i, j, \alpha} V_{i\,j} \left( \langle c_{i\, \alpha\, \downarrow} c_{j\, \alpha \, \uparrow}\rangle c_{i\, \alpha\, \downarrow}^{\dag} c_{j \, \alpha \, \uparrow}^{\dag} + H.c. \right)
\end{split}
\end{equation}
where $c_{i\, \alpha \, \sigma}^{\dag}$ and $c_{i\, \alpha \, \sigma}$ are respectively the creation and annihilation operators for an electron with spin $\sigma$ in the orbitals $\alpha = 1$ or $2$ on the $i$-th lattice site, $\mu$ is the chemical potential which is determined by the electron filling per site, corresponding to different doping. $t_{i\, j}^{\alpha \, \beta}$ are the hopping integrals. We choose the nonvanishing hopping elements as~\cite{YuanYenepl, HChen2013}: $ t^{\alpha\bar\alpha}_{\pm\hat x} =  t^{\alpha\bar\alpha}_{\pm\hat y}= t_1$, $ t^{11}_{\pm(\hat x+\hat y)} =  t^{22}_{\pm(\hat x-\hat y)}= t_2$, $ t^{11}_{\pm(\hat x-\hat y)} =  t^{22}_{\pm(\hat x+\hat y)}= t_3$, $ t^{\alpha\bar\alpha}_{\pm(\hat x\pm\hat y)}= t_4$, $ t^{\alpha\alpha}_{\pm\hat x} =  t^{\alpha\alpha}_{\pm\hat y}= t_5$, $ t^{\alpha\alpha}_{\pm2\hat x}= t^{\alpha\alpha}_{\pm2\hat y}= t_6$.

Besides, $V_{i\,j}\langle c_{i\, \alpha\, \downarrow} c_{j\, \alpha \, \uparrow}\rangle = \Delta_{i\, j}^{\alpha}$ is the SC bond pairing order parameter between site $i$ and $j$. Here, we only consider the next-nearest-neighbor (NNN) intraorbital pairing with strength $V_{i\,j} = V_{NNN} = V$, as a constant. This choice is consistent with the $s_{\pm}$ pairing~\cite{Ding2008, Mazin2008, FWang2009} and has been widely used in previous theoretical studies based on the BdG technique~\cite{Huang2011, Zhao2015}.

$H_{int}$ is the on-site interaction term. At the mean-field level, it can be written as
\begin{equation}
\begin{aligned}
\label{HamInt}
    H_{int} &= U \sum_{i, \alpha, \sigma \neq \bar{\sigma}} \langle \hat{n}_{i\, \alpha \, \bar{\sigma}} \rangle \, \hat{n}_{i\, \alpha \, \sigma} + U^{\prime} \sum_{i, \alpha \neq \beta, \sigma \neq \bar{\sigma}} \langle \hat{n}_{i\, \alpha \, \bar{\sigma}} \rangle \, \hat{n}_{i\, \beta \, \sigma} \\
    & + (U^{\prime} - J_{H}) \sum_{i, \alpha \neq \beta, \sigma} \langle \hat{n}_{i\, \alpha \, \sigma} \rangle \, \hat{n}_{i\, \beta \, \sigma}
\end{aligned}
\end{equation}
where $\hat{n}_{i\, \alpha \, \sigma} = c_{i\, \alpha \, \sigma}^{\dag} \, c_{i\, \alpha \, \sigma}$. The orbital rotation symmetry imposes the constraint $U^{\prime} = U - 2\,J_{H}$~\cite{constraint}.

The impurity part of the Hamiltonian $H_{imp}$ includes two parts, expressed as:
\begin{equation}\label{HamImp}
    H_{imp} = \sum_{I_{m}\, \sigma} V_{imp} c_{I_{m}\, \sigma}^{\dag} c_{I_{m}\, \sigma} + \sum_{I_{m}\, j\, \alpha\, \sigma} \left( \tilde{t}^{s\,\alpha}_{I_{m}\, j} c_{I_{m}\, \sigma}^{\dag} c_{j\, \alpha\, \sigma} + H.c. \right)
\end{equation}
where $c_{I_{m}\,\sigma}^{\dag}$ and $c_{I_{m}\,\sigma}$ are respectively the creation and annihilation operators for an electron with spin $\sigma$ on the $I_{m}$-th lattice site. The original Fe ion at $I_{m}$-th lattice site is substituted by a Zn ion. The first part of Eq.~\ref{HamImp} represents the on-site scattering. $V_{imp}$ is the scattering strength. Here $V_{imp} > 0$, which suggests that the Zn impurities behave like randomly distributed local-potential barriers imbedded in a sea of itinerant electrons.

\begin{figure}[htbp]
\centering
\subfigure[ ]{\includegraphics[width=\linewidth]{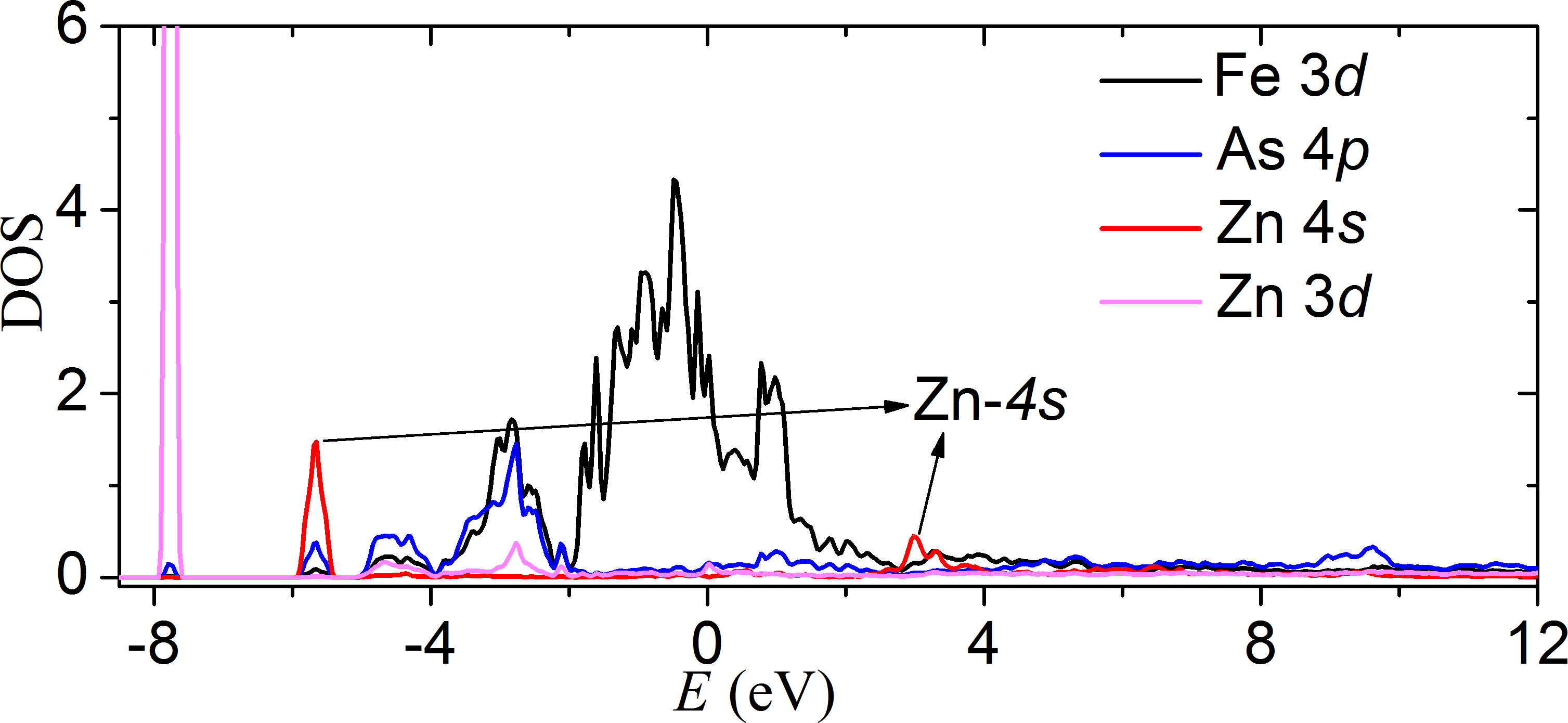}\label{pDOS}} \\
\subfigure[ ]{\includegraphics[width=0.32\linewidth]{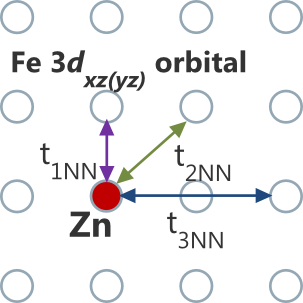}\label{hopping}}
\,\,\,\,
\subfigure[ ]{\includegraphics[width=0.62\linewidth]{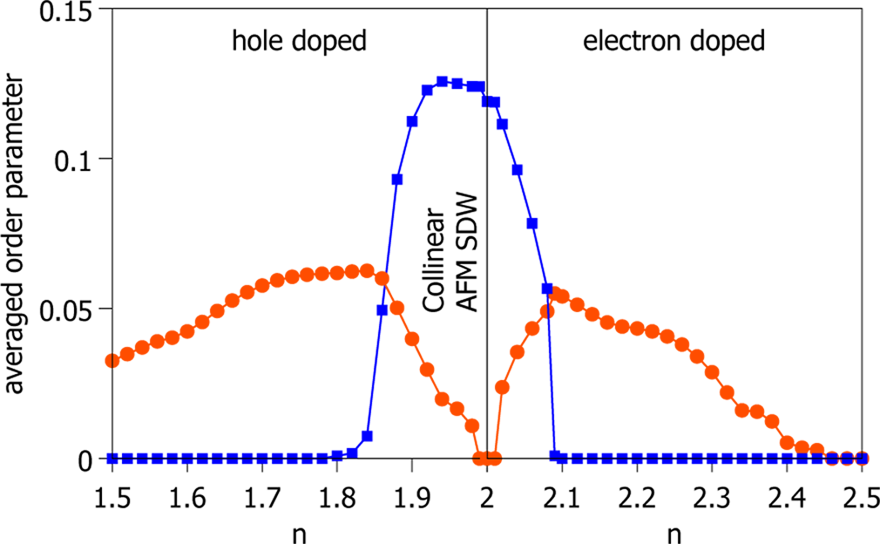}\label{totphase}}
\caption{(Color online) (a) Partial Fe-$3d$, Zn-$3d$, Zn-$4s$ and As-$4p$ density of states in substituted \BFA~\cite{footnote}. (b) The schematic diagram of the Zn-Fe hopping integrals between Zn-$4s$ orbital and Fe-$3d_{xz\,(yz)}$ orbital. The red solid filled circle represents the substituted-Zn ion and the gray circles represent the Fe ions. The purple, green, and blue solid line corresponds to the first-nearest-neighbor (1NN), second-nearest-neighbor (2NN) and third-nearest-neighbor (3NN) hopping terms, respectively. (c) Averaged magnetic and SC order parameters change under different doping level $n$ for \BFA compound.}
\end{figure}

Based on the spherical spatial orientation of the Zn-$4s$ orbital, the hopping integrals $\tilde{t}^{s\,\alpha}_{I_{m}\, j}$ between the $\alpha$-orbital of adjacent Fe ion at $j$-th lattice site and $s$-orbital of the substituted-Zn ion at $I_{m}$-th lattice site should be isotropic along different directions. As shown in Fig.~\ref{hopping}, we suppose the hopping terms between the substituted-Zn ion and adjacent Fe ions include first-nearest-neighbor (1NN) $t_{1NN}$, second-nearest-neighbor (2NN) $t_{2NN}$, and third-nearest-neighbor (3NN) $t_{3NN}$. For convenience, without introducing new parameters into our model, the strength of them are chosen to be: $t_{1NN} = t_{5}$, $t_{2NN} = t_{3}$, and $t_{3NN} = t_{6}$.

Now, we write down the matrix form of Eq.~\ref{HamTot} with basis $\psi_{i\alpha}=(c_{i\alpha\uparrow},c^\dagger_{i\alpha\downarrow})^{T}$, $H=\sum_{ij\alpha\beta}\psi^\dagger_{i\alpha}\, H_{BdG} \,\psi_{j\beta}$, and calculate the eigenvalues and eigenvectors of $H_{BdG}$:
\begin{equation}\label{BdG}
    \sum_{j\,\beta}
    \left(\begin{array}{cc}
        H_{i\, j\, \uparrow}^{\alpha\, \beta} & \Delta_{i\,j}^{\beta}\\
        \\
        \Delta_{i\,j}^{\beta *}& - H_{i\, j\, \downarrow}^{\alpha \, \beta}
    \end{array} \right)
    \left(\begin{array}{c}
        u_{j\, \beta}^{n} \\
        \\
        v_{j\, \beta}^{n}
    \end{array} \right)
    = E_{n}
    \left(\begin{array}{c}
        u_{i\, \alpha}^{n} \\
        \\
        v_{i\, \alpha}^{n}
    \end{array} \right)
\end{equation}
where, $H_{i\, j\, \sigma}^{\alpha\, \beta} \equiv [ H_{BCS} + H_{int} + H_{imp} ]_{ij\sigma}^{\alpha\beta}$, is the matrix-element for the single-particle Hamiltonian. And we have
\begin{eqnarray}
    \Delta_{i\,j}^{\alpha} &=& \frac{V_{ij}}{4}\sum_{n}( u_{i\, \alpha}^{n} v_{j\, \alpha}^{n*}
						+ u_{j\, \alpha}^{n} v_{i\, \alpha}^{n*})\,\mbox{tanh}\big(\frac{E_n}{2k_B T}\big) \label{locDel}\\
    \langle \hat{n}_{i\, \alpha \, \uparrow} \rangle &=& \sum_{n} \left| u_{i\, \alpha}^{n} \right|^{2} f(E_{n}) \nonumber \\
    \langle \hat{n}_{i\, \alpha \, \downarrow} \rangle &=& \sum_{n} \left| v_{i\, \alpha}^{n} \right|^{2} [1 - f(E_{n})]     \nonumber \\
    \langle \hat{n}_{i\, \alpha} \rangle &=& \langle \hat{n}_{i\, \alpha \, \uparrow} \rangle + \langle \hat{n}_{i\, \alpha \, \downarrow} \rangle \label{ni}
\end{eqnarray}
Here, $f(E_{n})$ is the Fermi-Dirac distribution function. To facilitate the discussion of physical quantities, we define the local magnetization and the $s_{\pm}$-wave projection of the SC order parameter at each site $i$, respectively as: $m_{i} = \frac{1}{4}\sum_{\alpha} (\langle \hat{n}_{i \alpha \uparrow} \rangle-\langle \hat{n}_{i \alpha \downarrow} \rangle)$, $\Delta_{i}= \frac{1}{8}\sum_{\delta, \alpha} \Delta_{i\,i+\delta}^{\alpha}$, where $\delta\in\{\pm\hat x\pm\hat y\}$. In addition, we also calculate the averaged values of $\langle |M|\rangle=\frac{1}{N}\sum_i |m_i|$ and $\langle \Delta_s\rangle=\frac{1}{N}\sum_i \Delta_i$, where $N$ is the number of Fe sites in the real-space lattice.

As mentioned before, substituted-Zn plays a heterovalent doping role into the system. The electron filling $n \equiv n_{Zn-free}$ represents the original level of a Zn-free system, it varies with the increase of substituted-Zn. For each constituent of \Zn, once the original doping level is determined, the total electrons in the system are initialized to the summation of the electrons ($n$) at Fe sites, and the electrons provided at the substituted-Zn ($4s^{2}$). In our self-consistent calculations of each specific doping level ($n$) and Zn concentration ratio ($x$), we set the total electron number of our system unchangeable, the chemical potential is determined by the total electron filling.

Throughout the paper, the energy is measured in unit of $t_{5}$. The temperature is set to be $T=0.0001$. Six hopping integrals are: $t_{1-6} = (0.09,\, 0.08,\, 1.35,\, -0.12,\, -1.00,\, 0.25)$. The on-site Coulomb interaction $U$ and Hund's coupling $J_{H}$ are set to be $3.5$ and $0.4$, respectively. The pairing strength $V = 1.3$. With these parameters, the dependence of the averaged magnetic and SC order parameters on the doping is illustrated in Fig.~\ref{totphase}, which is consistent with the experimental results~\cite{Pratt2009, Avci2012}. We choose the scattering strength of impurity $V_{imp} = 3$. The numerical calculations are performed on a $28\times28$ square lattice with periodic boundary conditions. In the multi-impurities cases, at each doping level $n$ and Zn-concentration value $x$, we calculate at least 20 different impurity configurations, in each of which substituted ions are distributed randomly. Furthermore, we try to avoid excessive concentration of impurities in a certain area. All the results we presented have been checked by using different initial values. Those results remain qualitatively similar, indicating the reliability of our calculations.

\section{Local electron density and local density of states (LDOS) around substituted-Zinc}
\begin{figure}[htbp]
\centering
\!\!\!\!
\subfigure[ $\,n = 1.80$, $x = 0.01$ ]{\includegraphics[width=0.45\linewidth, clip=true]{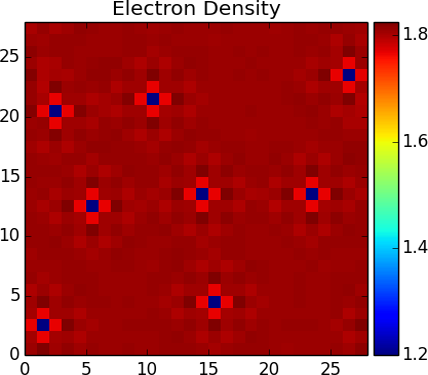}\label{180-8-den}}
\qquad
\subfigure[ $\,n = 1.80$, $x = 0.03$ ]{\includegraphics[width=0.45\linewidth, clip=true]{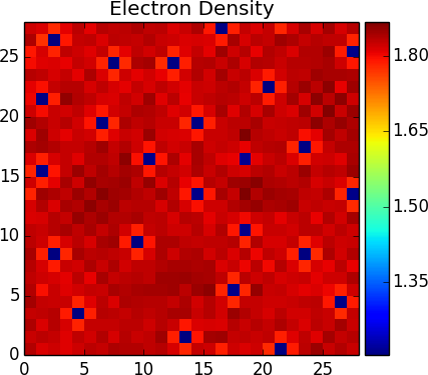}\label{180-24-den}}\\
\subfigure[ $\,n = 2.10$, $x = 0.01$ ]{\includegraphics[width=0.45\linewidth, clip=true]{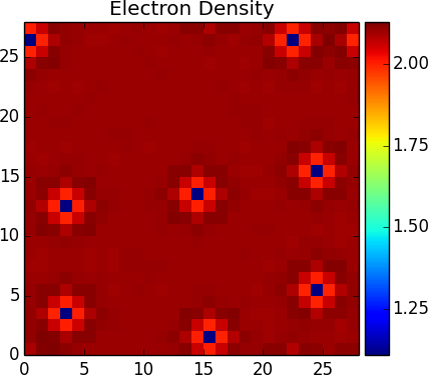}\label{210-8-den}}
\qquad
\subfigure[ $\,n = 2.10$, $x = 0.03$ ]{\includegraphics[width=0.45\linewidth, clip=true]{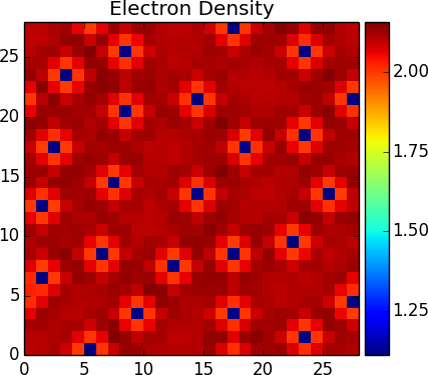}\label{210-24-den}}
\caption{(Color online) { Spatial profiles of local electron density at different doping level ($n$) and different Zn-concentration ration $x$. (a) $n = 1.80$, $x = 0.01$; (b) $n = 1.80$, $x = 0.03$; (c) $n = 2.10$, $x = 0.01$; (d) $n = 2.10$, $x = 0.03$. }
} \label{eleden}
\end{figure}

With $V_{imp} = 3 \sim 4$, the local electron density at the substituted-Zn ion is around 1. For convenience, we choose $V_{imp} = 3$ in our calculations. Fig.~\ref{eleden} shows the spatial profile of the local electron density under different conditions. At a specific doping level $n$, the local electron density around the substituted-Zn impurities just changes a little with different $x$.

\begin{figure}[htbp]
\centering
\!\!\!\!
\subfigure[ $\,n = 1.80$ ]{\includegraphics[width=0.45\linewidth, clip=true]{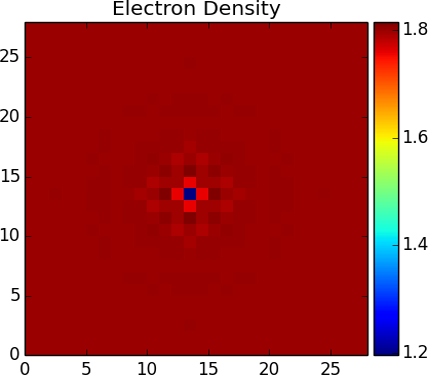}\label{180-1-Den}}
\qquad
\subfigure[ $\,n = 2.10$ ]{\includegraphics[width=0.45\linewidth, clip=true]{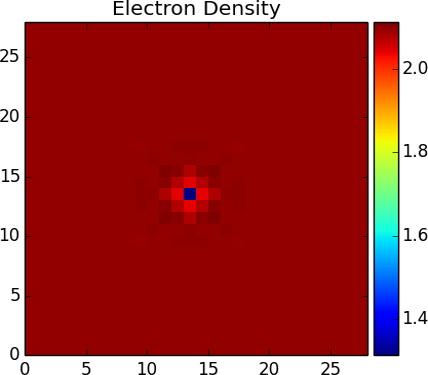}\label{210-1-Den}}
\\
\subfigure[ $\,n = 1.80$ ]{\includegraphics[width=0.48\linewidth, clip=true]{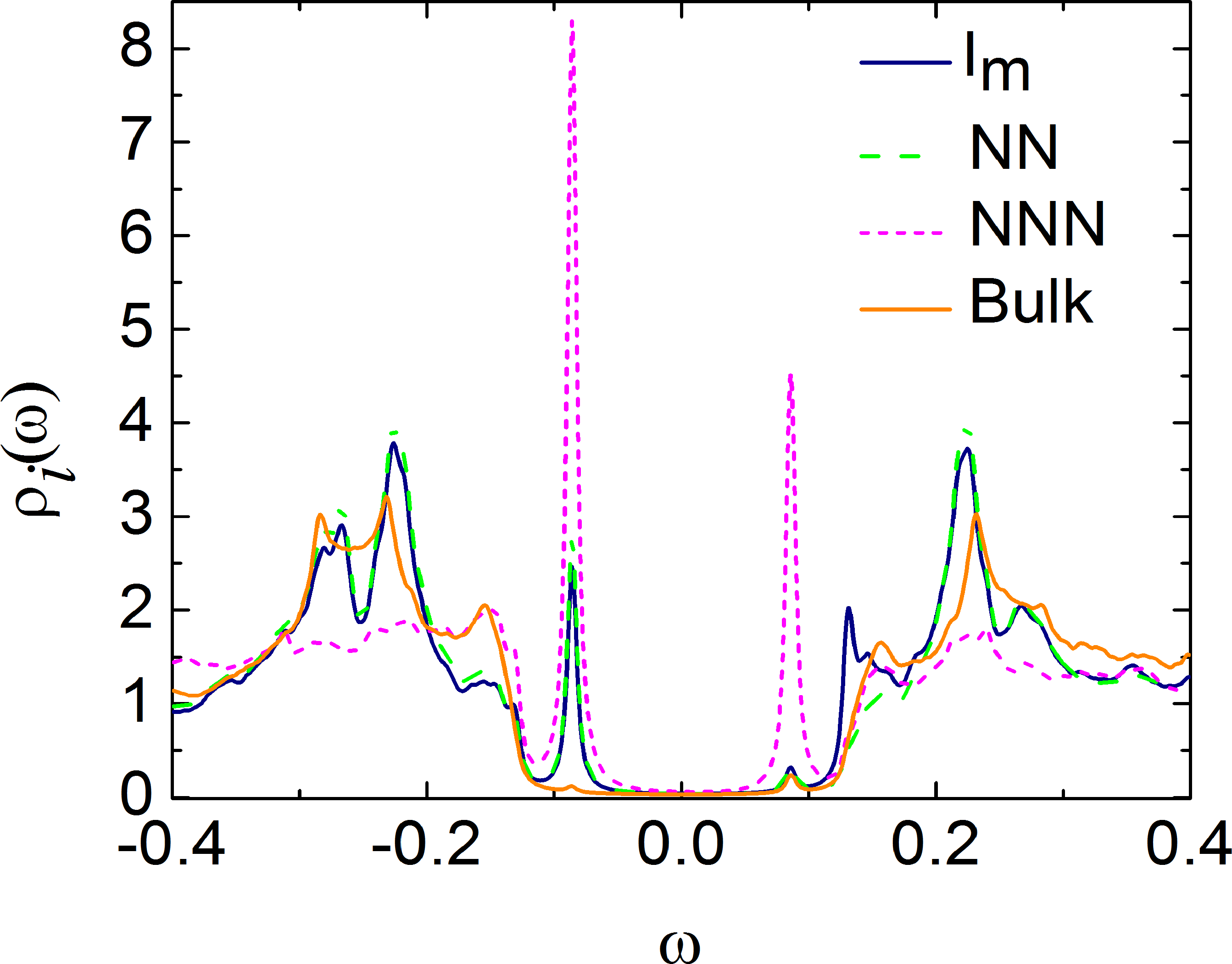}\label{180-14-14}}
\,
\subfigure[ $\,n = 2.10$ ]{\includegraphics[width=0.48\linewidth, clip=true]{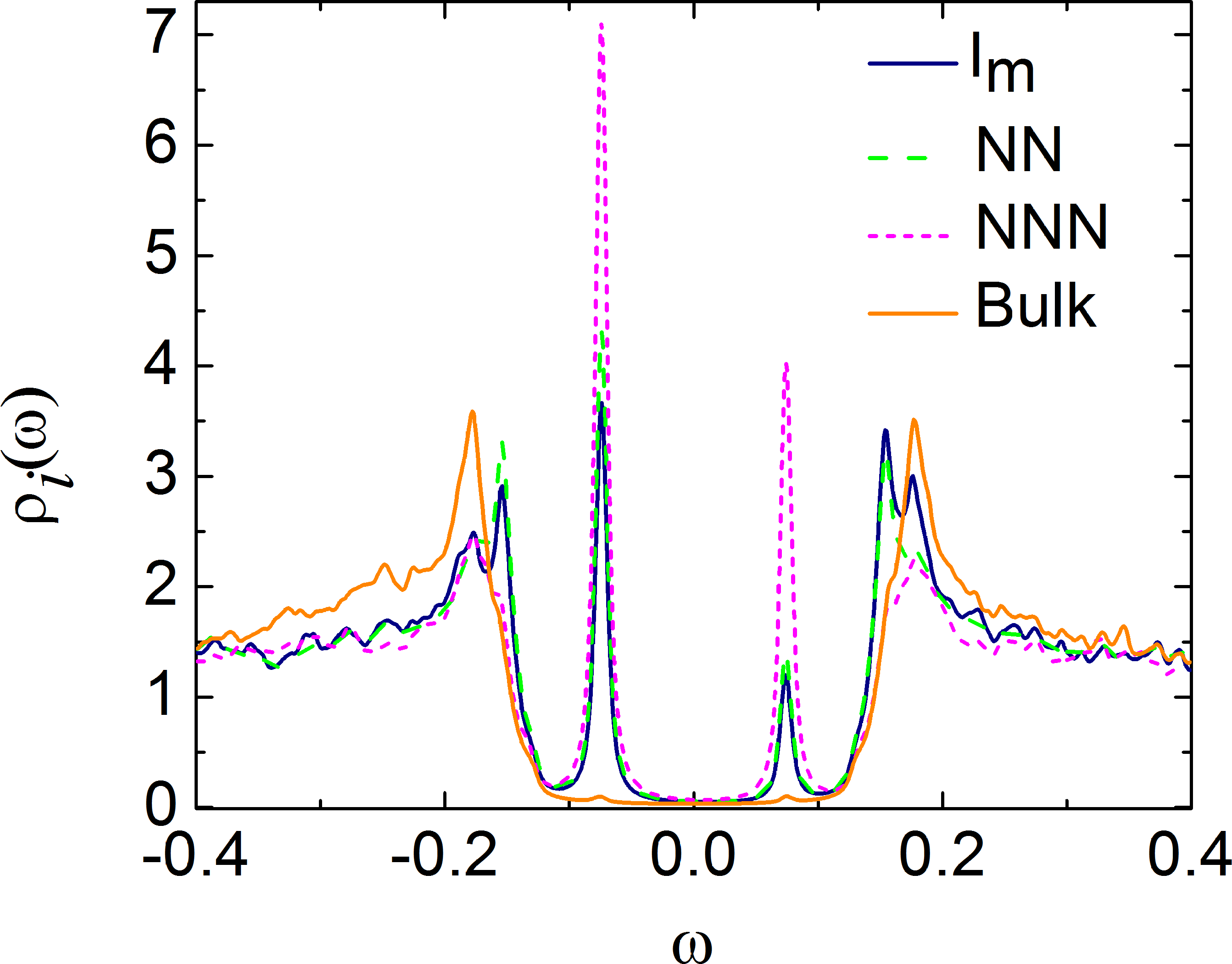}\label{210-14-14}}
\caption{(Color online) Spatial profiles of local electron density with single substituted-Zn ion at the center $I_{m} = (14, 14)$ at different doping level (a) $n = 1.80$, (b) $n = 2.10$. Corresponding LDOS spectra with (c) $n = 1.80$, (d) $n = 2.10$. The blue solid lines represent the LDOS at the impurity site. The green dash lines show the LDOS at the NN site of the impurity site. The magenta short dash lines show the LDOS at the NNN site of the impurity site. The orange solid lines show the LDOS at the site far away from the impurity.
} \label{LDOS}
\end{figure}

In order to further investigate the disorder effect of a single Zn impurity, we calculate the local density of states (LDOS) spectra near the impurity site. The LDOS can be expressed as
\begin{equation}\label{LDOSEq}
    \rho_{i}(\omega) = \sum_{n, \alpha} \left[ \big| u_{i\, \alpha}^{n} \big|^{2} \delta(E_{n} - \omega)
                        + \big| v_{i\, \alpha}^{n} \big|^{2} \delta(E_{n} + \omega) \right]
\end{equation}
where the delta function $\delta(y) = \Gamma/\pi(y^{2} + \Gamma^{2})$, and $\Gamma = 0.004$ is the quasiparticle damping. A $32 \times 32$ supercell is taken to calculate the LDOS. Two in-gap resonance peaks~\cite{Tao2011} emerge below and above the Fermi energy $E_f$ at the impurity site, nearest-neighbor (NN) sites, and NNN sites, as shown in Fig.~\ref{180-14-14} and Fig.~\ref{210-14-14}. The LDOS curves at these sites are similar. The intensity of the peak below $E_f$ is higher than that of the right peak above $E_f$. Moreover, the intensity of the peak below $E_f$ is the highest in NNN site, then lower in NN site, finally in the impurity site.

\section{SDW suppression with the existence of Zinc impurities}
\begin{figure}[htbp]
\centering
\!\!\!\!
\subfigure[ $\,n = 1.90$ ]{\includegraphics[width=0.45\linewidth, clip=true]{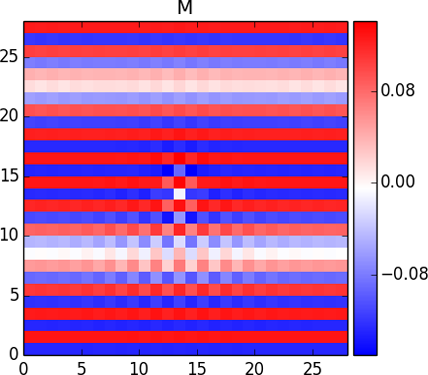}\label{190-1-mag}}
\qquad
\subfigure[ $\,n = 2.05$ ]{\includegraphics[width=0.45\linewidth, clip=true]{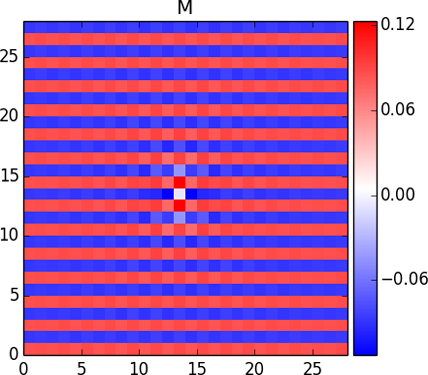}\label{205-1-mag}}\\
\subfigure[ $\,n = 1.90$ ]{\includegraphics[width=0.45\linewidth, clip=true]{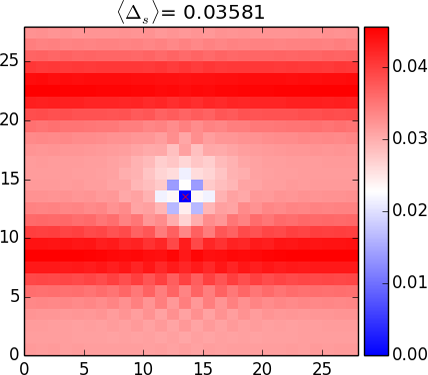}\label{190-1-sup}}
\qquad
\subfigure[ $\,n = 2.05$ ]{\includegraphics[width=0.45\linewidth, clip=true]{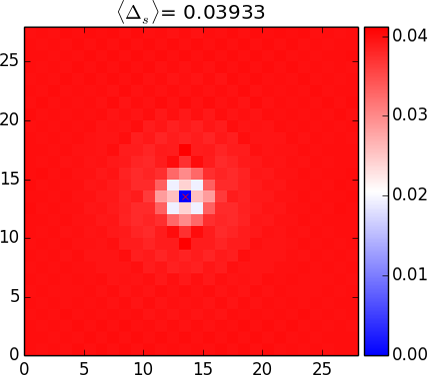}\label{205-1-sup}}
\caption{(Color online) Spatial profiles of local magnetic order parameter $M$ with single substituted-Zn ion at $I_{m} = (14, 14)$, at different doping level (a) $n = 1.90$, (b) $n = 2.05$. Spatial profiles of local superconductivity order parameter $\Delta_{s}$ with single substituted-Zn ion at $I_{m} = (14, 14)$, at different doping level (c) $n = 1.90$, (d) $n = 2.05$. $\langle \Delta_{s} \rangle$ represents the averaged value of $\Delta_{s}$.} \label{Single}
\end{figure}

We have studied the effect of substituted-Zn ions on the SDW order in doped \Zn. First, we focus on the single impurity effect on the underdoped systems. We put our single substituted-Zn at the center $I_{m} = (14, 14)$. According to the phase diagram (Fig.~\ref{totphase}), we choose the system at two different doping level $n = 1.90$ and $n = 2.05$, respectively. These two systems show the coexistence of collinear AFM SDW and SC. Fig.~\ref{190-1-mag} and Fig.~\ref{205-1-mag} shows the spatial profiles of local magnetic order with single substituted-Zn ion. Numerically in the system at $n = 1.90$, the AFM SDW is not so stable, but it is obvious. Both two cases give the similar results: no magnetic order at the substituted-Zn atom, while marginally suppressed magnetic order at the adjacent Fe sites.

\begin{figure}[htbp]
\centering
\!\!\!\!
\subfigure[ $\,x = 0.061$ ]{\includegraphics[width=0.45\linewidth, clip=true]{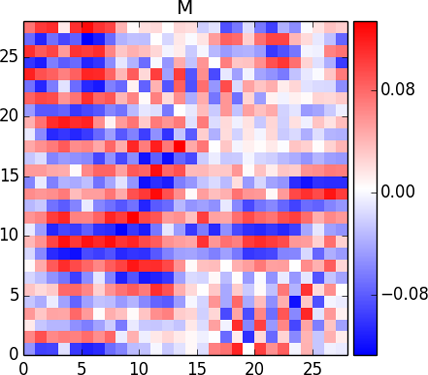}\label{2-48-mag}}
\qquad
\subfigure[ $\,x = 0.25$ ]{\includegraphics[width=0.45\linewidth, clip=true]{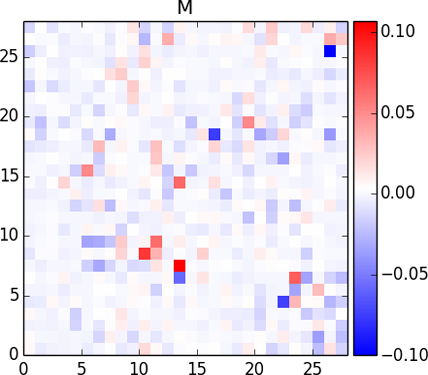}\label{2-196-mag}}
\caption{(Color online) Spatial profiles of local magnetic order parameter $M$ at doping level $n = 2.00$, with different concentration of substituted-Zn ions: (a) $x = 0.061$, (b) $x = 0.25$. } \label{2Mag}
\end{figure}

According to the suppression of AFM SDW in the system with single impurity, we investigate the multi-impurities cases.
Previous discussions show that substituted-Zn in \Zn system provides electrons into the system. Thus the effect on the AFM SDW of multi-substituted-Zn system comes from the scattering of substituted-Zn, meanwhile, the effect also comes from the change of doping level in the system. The SDW in the hole-doped region is complicated. It will be suppressed until the concentration of substituted-Zn ions is a relatively large value in the system. Correspondingly, we choose $n = 2.00$ to study the SDW order. Fig.~\ref{2Mag} shows the spatial profiles of local magnetic order parameter with different Zn-concentration ratio $x$.
With $x = 0.061$, the strength of local magnetic order parameter is suppressed in the occupied substituted-Zn ions areas, while in the few substituted-Zn ions areas, the collinear AFM SDW is still distinct. Although in our discussions, the interaction between Zn impurities are neglected while we avoid excessive concentration of impurities in an area, we still calculate the case with $x = 0.25$. As shown in Fig.~\ref{2-196-mag}, the SDW is completely destroyed, which is consistent with the experimental results~\cite{Ideta2013}.

\section{Superconductivity suppression with the increase of Zinc impurities}
In the following, we study the effect of substituted-Zn ions on the SC order parameter at different doping levels. First, we focus on single substituted-Zn ion at the center of lattice site $I_{m} = (14, 14)$ at two different doping levels: $n = 1.90$ and $n = 2.05$. Fig.~\ref{190-1-sup} and Fig.~\ref{205-1-sup} illustrate the spatial profiles of local SC order parameter in each case, respectively. It is very clear that the SC order parameter on the adjacent sites is suppressed around the substituted-Zn ion.

\begin{figure}[htbp]
\centering
\subfigure[ $\,n = 2.05, \, x = 0.01$ ]{\includegraphics[width=0.45\linewidth, clip=true]{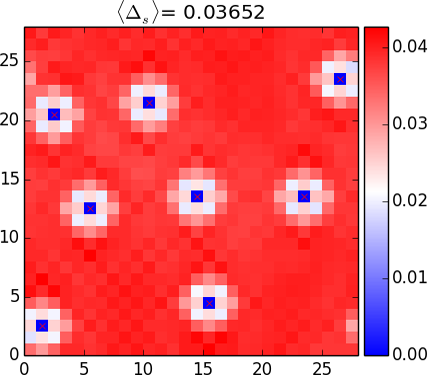}\label{205-8-Sup}}
\qquad
\subfigure[ $\,n = 2.10, \, x = 0.01$ ]{\includegraphics[width=0.45\linewidth, clip=true]{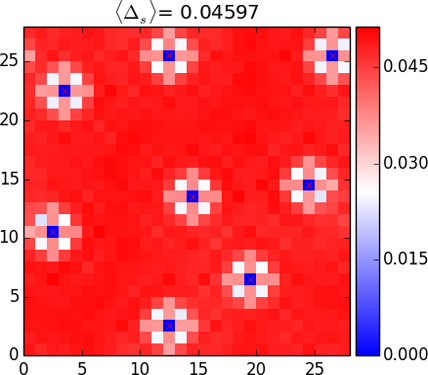}\label{210-8-Sup}}
\\
\subfigure[ $\,n = 2.05, \, x = 0.061$ ]{\includegraphics[width=0.45\linewidth, clip=true]{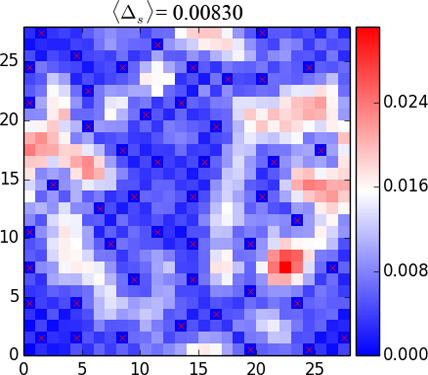}\label{205-48-Sup}}
\qquad
\subfigure[ $\,n = 2.10, \, x = 0.061$ ]{\includegraphics[width=0.45\linewidth, clip=true]{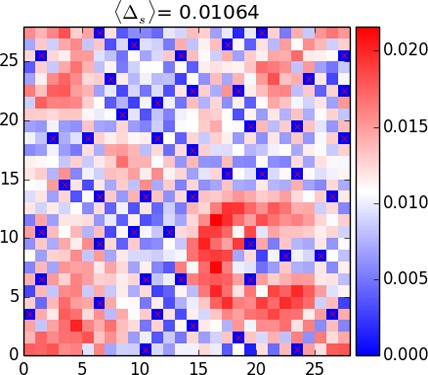}\label{210-48-Sup}}
\\
\subfigure[ $\,n = 1.80, \, x = 0.01$ ]{\includegraphics[width=0.45\linewidth, clip=true]{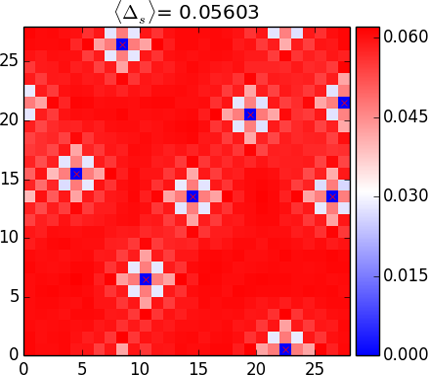}\label{180-8-Sup}}
\qquad
\subfigure[ $\,n = 1.90, \, x = 0.01$ ]{\includegraphics[width=0.45\linewidth, clip=true]{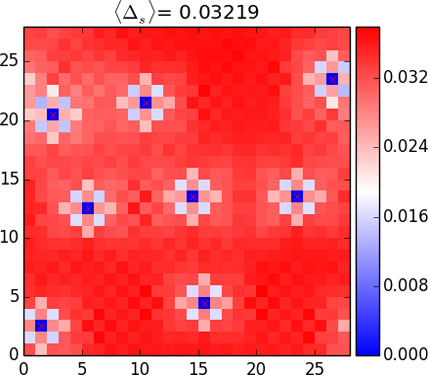}\label{190-8-Sup}}
\\
\subfigure[ $\,n = 1.80, \, x = 0.03$ ]{\includegraphics[width=0.45\linewidth, clip=true]{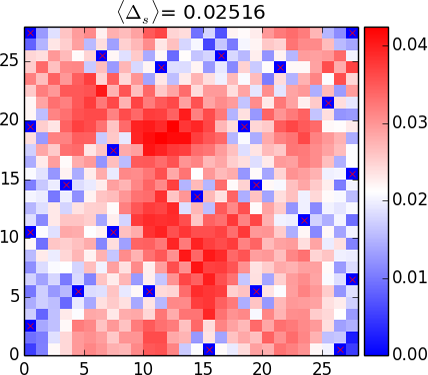}\label{180-24-Sup}}
\qquad
\subfigure[ $\,n = 1.90, \, x = 0.03$ ]{\includegraphics[width=0.45\linewidth, clip=true]{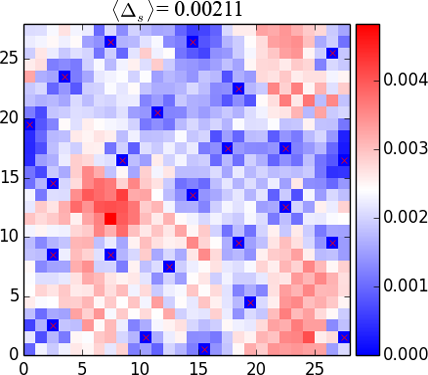}\label{190-24-Sup}}
\caption{(Color online) Spatial profiles of local SC order parameter $\Delta_{s}$ with different conditions. $\times$ (red) represents the position of the substituted-Zn ions. $\langle \Delta_{s} \rangle$ represents the averaged value of $\Delta_{s}$.} \label{MultiSup}
\end{figure}

We investigate the effect of multi-Zn impurities on SC with different $x$ at the electron-doped region and hole-doped region, respectively. According to Fig.~\ref{MultiSup}, the SC is suppressed because of the substituted-Zn ions in \Zn system regardless of the doping level. In the electron-doped region (Fig.~\ref{205-8-Sup} - Fig.~\ref{210-48-Sup}), with $x = 0.01$, our results show the SC order is suppressed around the substituted-Zn ions. While with $x = 0.061$, the SC order becomes localized, and the high intensity spots of the SC order parameter appear most likely in the Zn-free areas. The similar results are obtained in the hole-doped region (Fig.~\ref{180-8-Sup} - Fig.~\ref{190-24-Sup}). However, the SC is suppressed severely in the hole-doped region. At $n = 1.90$ and with $x = 0.03$, the SC order is nearly destroyed.

\begin{figure}[htbp]
\centering
\!\!\!\!
\subfigure[ \, electron-doped region ]{\includegraphics[width=0.48\linewidth, clip=true]{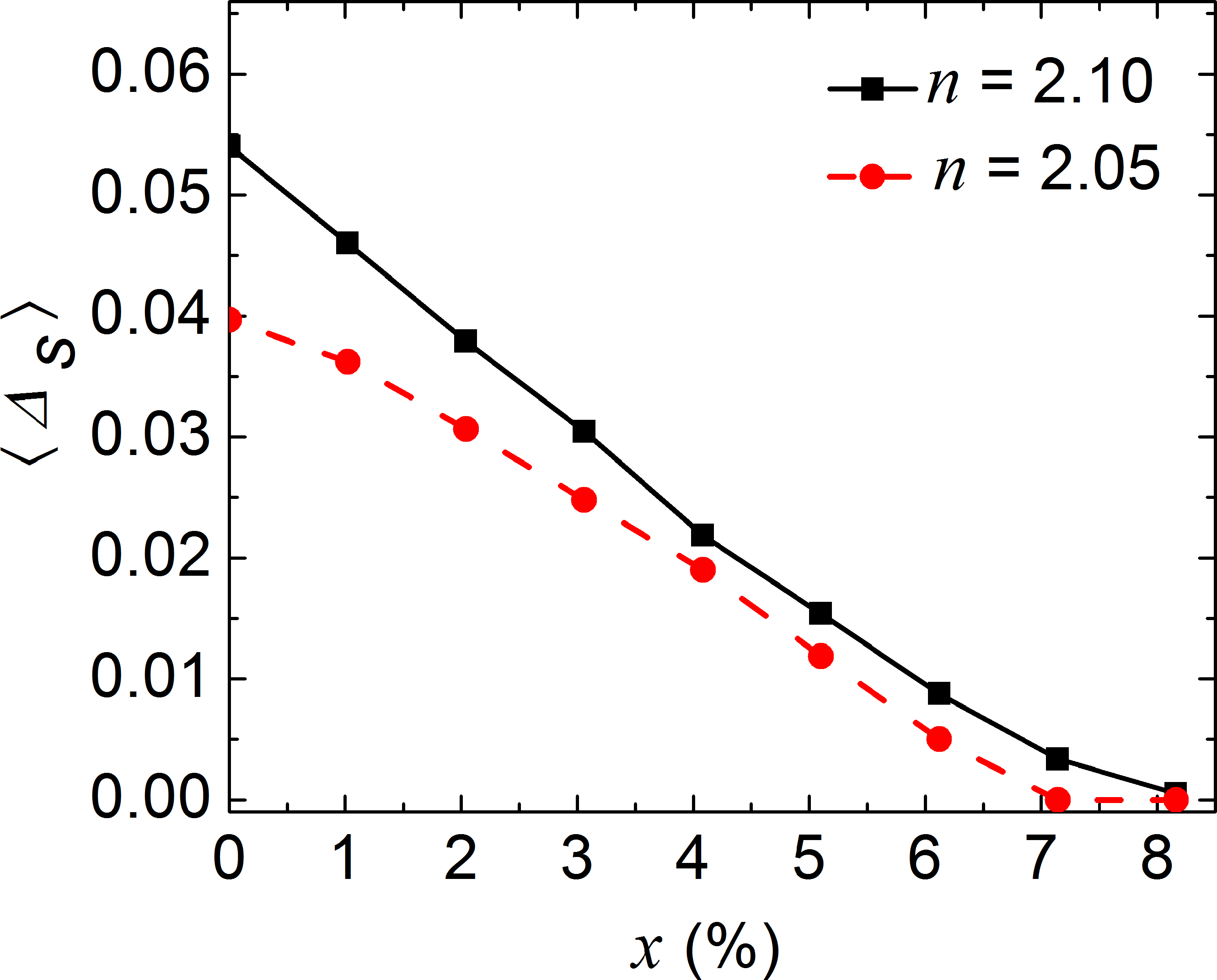}\label{electron}}
\,
\subfigure[ \, hole-doped region ]{\includegraphics[width=0.48\linewidth, clip=true]{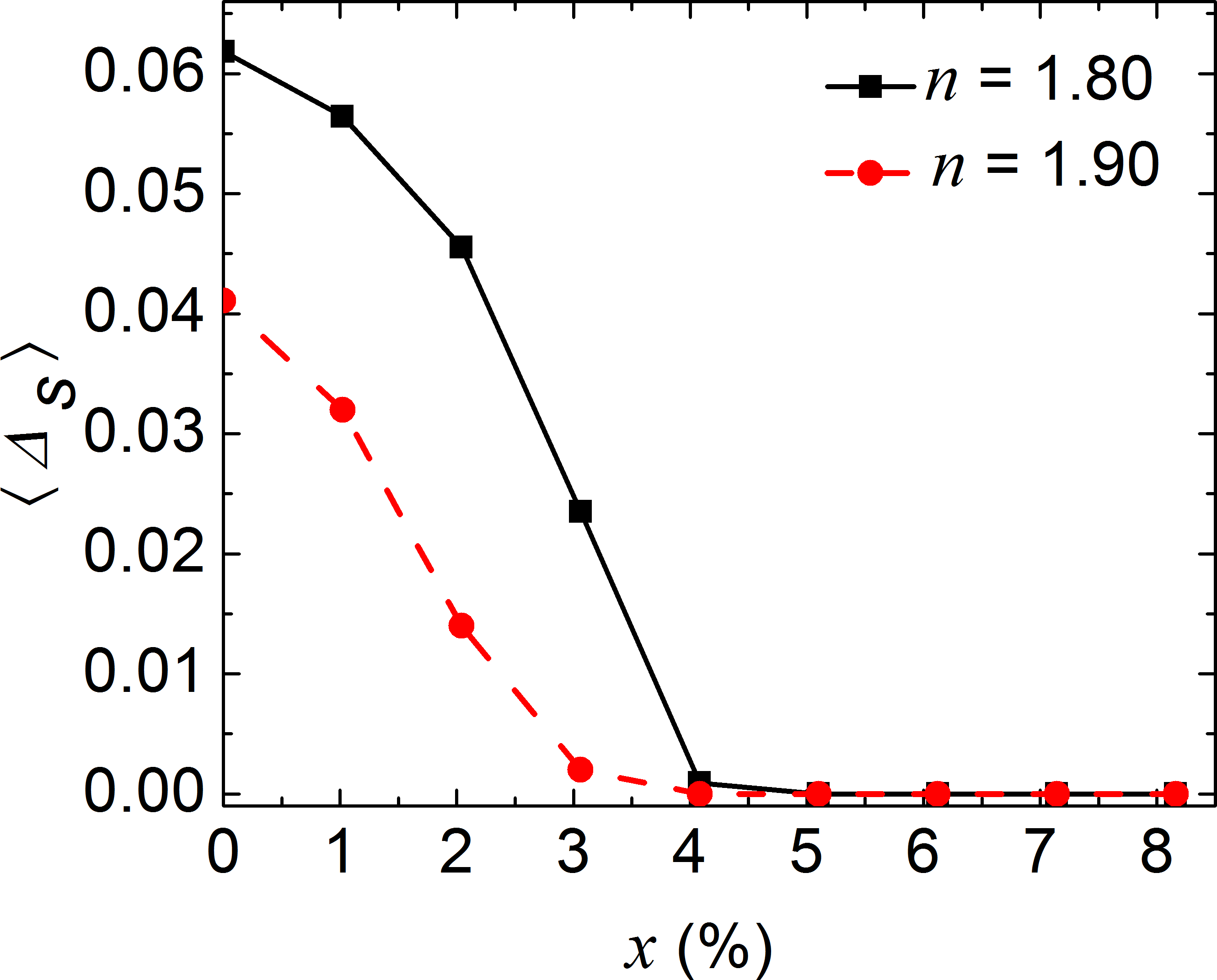}\label{hole}}
\caption{(Color online) Averaged SC order vs. the ratio of substituted-Zn ions in \Zn system in the (a) electron-doped region, and (b) hole-doped region.} \label{trend}
\end{figure}

Furthermore, we averaged over 20 different impurity configurations for each doping level $n$ and the Zn-concentration ratio $x$. Then we obtain the dependence of the averaged SC order parameters on different $x$ in \Zn system at a certain doping level $n$, as shown in Fig.~\ref{trend}. Fig.\ref{electron} illustrates the relation in the electron-doped region. SC order linearly decreases with increasing substituted-Zn ions in \Zn system. Around $x=0.08$, the SC roughly disappears, no matter the doping level is in the optimal-doped ($n = 2.10$), or the under-doped ($n = 2.05$) region. These results agree well with the experiments~\cite{Li2011B, Li2012}. Fig.\ref{hole} shows the relation in the hole-doped region. The SC order decreasing trend is different from that in the electron-doped region. The SC is suppressed more severely, then destroyed with a smaller Zn-concentration ratio $x$ in the system. Our calculations give comparable results with recent experiments~\cite{Li2015}, which show that SC in the hole over-doped \BFA almost disappear around $x = 0.03$.

\section{conclusion and discussions}
The conventional understanding of the substituted-Zn in the \BFA assumes that Zn plays an isovalent role, and it does not provide any carriers into the system. With this assumption, even in the electron under-doped system with $n = 2.05$, the SC order appears to drop much faster as the increase of substituted-Zn, comparing with the experimental measurements~\cite{Li2012}. This motivates us to reexamine the nature of Zn in the doped \BFA. Although there exists no experiment showing how large the valence of the Zn impurity should be, a recent DFT work~\cite{Khan2014}  indicates that Zn provides effective additional electron into the system. We believe the additional electron should not come from Zn-$3d$ orbitals. Thus we perform the first-principles calculations and resolve the projected partial Fe-$3d$, Fe-$4s$, Zn-$3d$, and Zn-$4s$ density of states in the substituted-Zn \BFA system. Our results indicate that the Zn-$3d$ orbitals are far below the Fermi energy, while the Zn-$4s$ level is narrowly distributing above and below the Fermi energy, and more important is that the empty Zn-$4s$ energy is not far above the Fermi energy. Thus the substituted-Zn should not be isovalent that is different from the conventional understanding. For the sake of convenience, we choose the proper local band parameters to make the valence of Zn close to 1. With this assumption, we are able to account for the suppressions of SC by Zn impurities in both electron- and hole- doped \BFA with and without the SDW order as measured by experiments ~\cite{Li2011B, Li2012, Li2015} using the same set of local band parameters.

\section{Acknowledgement}
The authors thank J. X. Zhu, Y.-Y. Tai, L. H. Pan, J. Kang for useful comments and discussions. This work was supported by the Texas Center for Superconductivity at the University of Houston, the Robert A. Welch Foundation under Grant No. E-1146 (Y. Y. Zhao and C. S. Ting), and the National Science Foundation under Grant No. DMR-1206839 and DMR-1507371 (B. Li and K. E. Bassler). W. Li is supported by the Strategic Priority Research Program (B) of the Chinese Academy of Sciences (Grant No. XDB04040300), the National Natural Science Foundation of China (Grants No. 11227902 and No. 11404359), and the Shanghai Yang-Fan Program (Grants No. 14YF1407100). H.-Y. C. is supported by Ministry of Science and Technology of Taiwan under Grant No. 104-2112-M-003-003-MY3 and 102-2112-M-002-003-MY3, and National Center for Theoretical Science of Taiwan.


\begin{references}
\bibitem{footnote} The first-principles calculations presented in this work were performed using the all-electron full potential linear augmented plane wave plus local orbitals method~\cite{method} as implemented in the WIEN2K code~\cite{Wien2k}. The exchange-correlation potential was calculated using the generalized gradient approximation as proposed by Pedrew, Burke, and Ernzerhof~\cite{GGA}. These calculations were perform using the optimized $2\times 2$ supercell structure with one Fe atom substituted by Zn.
\end{references}
\end{document}